\documentclass{iau}

\usepackage{amsmath}
\usepackage{graphicx}
\usepackage{multirow}

\begin{document}

\lefttitle{Dionysios Gakis}
\righttitle{Indigenous narratives into modern stellar astrophysics}

\jnlPage{1}{7}
\jnlDoiYr{2026}
\doival{10.1017/xxxxx}

\aopheadtitle{Proceedings IAU Symposium}
\editors{D.W. Hamacher \&  J. Holbrook, eds.}

\title{Echoes of Star Stories: Integrating Indigenous narratives into modern stellar astrophysics}

\author{Dionysios Gakis$^1$}
\affiliation{$^1$Department of Astronomy, The University of Texas at Austin, \\ 2515 Speedway Blvd., Austin, TX 78712, USA}

\begin{abstract}
Indigenous narratives have long preserved observations of celestial phenomena, offering insights that resonate with modern astrophysical research. Deeply embedded in cultural traditions, these stories describe events such as stellar variability, supernovae, eclipses, and planetary alignments. Indigenous communities have also used the stars for navigation and calendar systems, reflecting a sophisticated understanding of celestial patterns. These narratives not only complement historical records but offer human-centric perspectives on stellar life cycles that often parallel modern models. Drawing from traditions from diverse Indigenous communities — including Aboriginal Australians, Pueblo peoples, Inuit, and Polynesians — this paper highlights the profound connections between Indigenous knowledge and stellar astrophysics. Integrating these traditions acknowledges their value, safeguards their relevance in the Space Age, and fosters mutual learning through cross-cultural collaboration. We explore how such narratives can enrich education, public engagement, and even inspire scientific hypotheses, bridging cultural heritage and modern astrophysics.
\end{abstract}


\maketitle

\section{Introduction}

Indigenous knowledge systems around the world have long incorporated stellar phenomena into their cosmologies, oral traditions, and daily practices. Indigenous narratives preserve more than cultural meaning; they also contain empirical evidence and qualitative records of stellar phenomena such as supernovae and transient eruptions, stellar variability, obscured star birth regions, metaphors of stellar evolution, and multiple star associations.

Indigenous narratives are often empirical, reflecting precise and long-term astronomical observations. Their cultural framing is not a limitation, but a way of preserving and transmitting knowledge. These stories serve as mnemonics, ethical guides, seasonal indicators, and in some cases mirror modern astrophysics. Hence, oral traditions are valuable from both a historical and scientific perspective, bridging Indigenous knowledge with modern science.

While this paper explores Indigenous knowledge related to stellar phenomena, it is important to recognize that Indigenous peoples have engaged with the night sky in far more diverse ways. Their astronomical knowledge extends to planetary movements, lunar phases, tidal patterns, and the use of celestial bodies for navigation and calendars. Examples include the Moon and Venus for timekeeping, the Pleiades and Vega for marking seasonal change, and the alignment of stone arrangements with solstitial points and cardinal directions \citep[e.g.,][]{2013JAHH...16..325H,2016PASA...33...39N,2018iau3.book...17H,wade2019polynesian}. These practices reflect a deep and empirical understanding of the cosmos beyond stellar events alone.

This paper is organized in two main parts. Section \ref{sec:sec2} presents examples of stellar phenomena as recorded and interpreted within Indigenous knowledge systems. Section \ref{sec:sec3} focuses on the integration of these traditions into modern science, with emphasis on collaboration, education, and the relevance of Indigenous knowledge in contemporary contexts.

\section{Indigenous Perspectives on Stellar Phenomena} \label{sec:sec2}

\subsection{Explosions in the Sky: Novae, supernovae and transient events}


Novae and supernovae represent some of the most dramatic events in stellar evolution. 
These events, though rare, are bright enough to remain in cultural memory. Narratives and material culture from Indigenous communities across the globe have been interpreted as possible records of such transient phenomena. While confirming these associations is challenging—due to ambiguities in dating, sky location, and symbolic representation—one case stands out.

The Great Eruption of $\eta$ Carinae (1837–1858), a luminous blue variable that briefly became the second-brightest star in the sky, is the only known case with confirmed Indigenous documentation. The Boorong people of northwestern Victoria incorporated this “new red star” into their oral traditions, naming it Collowgullouric War, the wife of War (Canopus). Stanbridge’s notes \citep{stanbridge1857astronomy} specify its brightness, redness, and position in Robur Carolinum, aligned with $\eta$ Carinae’s peak \citep{2010JAHH...13..220H,2011MNRAS.415.2009S}. This record shows how oral traditions can capture real-time stellar transients and preserve them with remarkable reliability. 

Several other Indigenous accounts have been proposed as possible records of historical supernovae, though none meet the full set of criteria outlined by \cite{2014JAHH...17..161H}—including explicit mention of a new star, known location and timing, and an identifiable remnant.

A well-known pictograph at Chaco Canyon, New Mexico, shows a large star beside a crescent moon. Some interpret this as the 1054 supernova, documented in Chinese records, but the lack of matching oral tradition and symbolic ambiguity render the case inconclusive \citep{brandt19791054,2018JAHH...21....7S}. A concentric-circle petroglyph in Arizona has been linked to SN 1006, but again, no corroborating oral context exists \citep{2006S&T...112R..17C}. A story from Yolngu at Arnhem Land describes a new star appearing after the deaths of two brothers, later forming a triangular star group. Although evocative, the account is mythic and lacks a clear date \citep{2014JAHH...17..161H}. Petroglyphs by Tiwanaku in the Andes dated to $\sim$1000 CE have been interpreted as recording a stellar explosion, but remain speculative \citep{2002JAVSO..31...54T}. A Chamorro narrative from Guam describes a “new star” that appeared brightly in the sky before vanishing, which may refer to SN 1054, though it lacks details such as sky location or precise dating \citep{villaverde2000camel}.

A particularly intriguing case is the Māori concept of \textit{Mahutonga}, the “star of the south that remains invisible.” Some researchers propose that this may preserve the memory of a southern supernova—possibly SN 185 near $\alpha$ Centauri or the 13th-century supernova remnant RX J0852.0–4622 \citep{best1922astronomical,2004Arch...18..110G,wade2019polynesian}. While speculative, the consistency in themes across Polynesian and other global traditions—such as stars that appear, guide migration, and disappear—offers a compelling example of oral memory spanning centuries.


\subsection{Flickering fires: Stellar variability}


While not as dramatic as supernovae, stellar variability has also been recorded in Indigenous traditions with remarkable precision. Several Aboriginal oral traditions accurately describe the brightness changes of red giant and supergiant stars, centuries before Western science formally recognized their variability. In the Kokatha story of Nyeeruna, Betelgeuse is a hunter whose “fire magic” intensifies during pursuit and fades with rejection, reflecting its semiregular pulsation between magnitude 1.3 and 0.0. Aldebaran, as the foot of a protective sister, also brightens in defiance, encoding smaller but still visible variability \citep{2018AuJAn..29...89H,2008MsT..........2F}.

In the Waiyungari tradition, Antares becomes a young man whose brightening marks sexual desire, surrounded by two companions in the Scorpius constellation \citep{clarke1999waiyungari}. This reflects Antares’s variation over a 4.5-year cycle from magnitude 1.6 to 0.6, linking its visibility with ceremonial timing \citep{2018AuJAn..29...89H}. Though initially misidentified as Mars, cross-checked narratives confirm it as a stellar account—offering the only known Indigenous records of naked-eye stellar pulsation \citep{2018AuJAn..29...89H,2018JAHH...21....7S}.

\subsection{Shapes in the shadows: Dark clouds and star birth regions}

While some traditions recorded stars that change over time, others focused on the shadows of the Milky Way, finding meaning not in starlight but in its absence \citep{2020JAHH...23..390G}. Aboriginal Australians in particular identified dark regions of the Milky Way as vivid celestial figures. The most widespread is the “Emu in the Sky,” traced not by stars but by the dark lanes of interstellar dust. Its head lies in the Coalsack Nebula near the Southern Cross, with a body stretching through Scorpius and Sagittarius. Its seasonal orientation was used to time emu nesting and egg collection, embedding ecological knowledge in the sky \citep{2013arXiv1311.0076F}. In Noongar culture, the Emu’s rise signals the onset of mokkar—the winter rains. 

Other dark cloud forms appear across the continent. Among the Wiradjuri, a cosmic serpent slithers between the Southern Cross and Vela (formed from the Coalsack and the Milky Way dark rift). In the Kimberley and Euahlayi traditions, beings like kangaroos (Bandaar) and crocodiles (Garriyas) are mapped to dark spaces. These narratives show that Indigenous astronomers paid attention not only to light, but also to its absence—recognizing the Milky Way’s obscuring nebulae as key celestial features \citep{1998nsaa.book.....J,2009edia.book.....N,2016PASA...33...39N}.

Similar interpretations exist far beyond Australia. In the Andes, the Inca saw dark shapes in the Milky Way—llamas, serpents, foxes—traveling across a celestial river (Mayu), with each animal linked to seasonal rhythms and fertility. The llama constellation (Yacana), with eyes marked by Alpha and Beta Centauri, was accompanied by a nursing offspring, aligning with the birthing season of terrestrial llamas. The Milky Way’s patchy structure was understood as a space of “life,” where animals moved through darkness rather than light \citep{1981aces.book.....U}. These constellations were spiritually alive, shaping agriculture and social order, a perspective also documented in the South American lowlands \citep[e.g.,][]{2023Land...12..805G}.

In the Amazon, Indigenous groups such as the Bororo and Tupi also mapped animals into the dark rift, with figures like the rhea or peccary stretching through the galactic bulge \citep{2020JAHH...23..390G}. Among the Maya, the same dark band was seen as Xibalbá be—the road to the underworld—underscoring its spiritual and cosmological weight \citep{schele1990forest}.

\subsection{Cosmic life cycles: Death, rebirth, and stellar evolution}

Across many Indigenous traditions, stars are not simply distant objects, but living participants in cycles of creation, transformation, and return. In Aboriginal Australian cultures, stars often represent ancestral spirits. When someone dies, a new star may appear, symbolizing their journey to the sky. The Yolngu, for example, tell of two drowned brothers who become stars, their brightness and position reflecting kinship and fate \citep{2014JAHH...17..161H,wells1973stars}. Though modern astronomy explains such appearances with novae or red giants, these traditions focus on spiritual, not physical, transitions, and no specific related astronomical event has been confirmed.

In Inuit cosmology, stars are the souls of people and animals, forming a celestial community rich with personal narratives. Single stars represent individual beings, as it was believed that each soul becomes a single star; in contrast, inanimate objects are depicted as star clusters or groups \citep{1998asia.book.....M}. This loosely parallels modern understanding that most stars form in groups and later drift apart.

Andean and Amazonian narratives also link stars with life and death. The Inca identified animals in the Milky Way’s dark dust clouds, whose risings and settings marked seasonal cycles of growth and fertility \citep{1981aces.book.....U}. These clouds are now known to be cold molecular regions—stellar nurseries where stars are born.

A vivid Arawak–Inca myth tells of a son of the Sun who becomes a star, is swallowed by the serpent of Scorpius, and later reemerges as Antares—symbolizing passage through cosmic death and return \citep{1981aces.book.....U}.  Astronomically, Antares is a red supergiant in a late stage of stellar evolution and is expected eventually to end as a supernova. The myth, however, frames this kind of cosmic death as re-emergence and renewal.

In Māori tradition, stars were placed in the sky by the god Tāne after separating earth and sky, suggesting a divine origin for the stars \citep{best1922astronomical}. This narrative links star formation to divine order, suggesting stars had an origin, not eternal existence.

Pueblo peoples, including the Navajo and Tewa, also embed stars in moral and cosmic narratives. In one Navajo story, Black God arranges stars with order, including placing the Pleiades on his temple. But Coyote disrupts this order by scattering stars chaotically, introducing disorder into the cosmos \citep{haile1947starlore,1983Arch....6...48C,newcomb1990navaho}. These tales reflect the tension between cosmic order and the seeming randomness of the stars.

\subsection{Celestial relationships: Binary stars and stellar clusters}

Many Indigenous traditions include detailed observations of stellar groupings, ranging from tightly bound pairs to dense clusters. The Boorong of western Victoria, for example, recognized the close pairing of $\alpha^1$ and $\alpha^2$ Cap—stars separated by just $0.1^\circ$—as a single ancestral figure named Collenbitchick \citep{2010JAHH...13..220H,stanbridge1857astronomy}, even though it is known today that they are an optical double rather than a physical binary. 
Such an association reflects careful naked-eye observation of close stellar pairs and the subtle distinctions between single and binary systems.

While such examples center on visual binaries, other traditions hint at more complex stellar systems. Among the Dogon of Mali, accounts describe Po Tolo—a dense, invisible companion to Sirius—along with Emme Ya Tolo and other unseen bodies, forming what some sources describe as a four-star system. Po Tolo has been interpreted as a reference to Sirius B, the white dwarf companion to Sirius A discovered telescopically in the 19th century \citep{griaule1950systeme,2007siri.book.....H}. The claim was initially debated in the literature, but is now generally interpreted as reflecting later influence from Western astronomy \citep{van1991dogon}. Regardless, the story engages with ideas of multiple stars and invisible companions, both of which are important in modern astrophysics.

Beyond binary stars, Indigenous knowledge systems often describe star clusters in relational or narrative terms. Notably the Pleiades 
are widely recognized across the globe. Variations in their brightness or visibility—particularly their heliacal rising—feature in stories that signal seasonal change or social rituals. Such long-term attentiveness may encode subtle shifts in visibility due to atmospheric or environmental factors, echoing modern studies of stellar variability and sky transparency.

\section{Integration, Collaboration, and Public Engagement} \label{sec:sec3}

These Indigenous perspectives reveal a long-standing engagement with stellar phenomena, linking observation, memory, and meaning, offering qualitative records of astrophysical events. Integrating these cultural viewpoints not only enhances scientific knowledge but affirms the value of Indigenous worldviews in interpreting our shared universe.

The Indigenous narratives explored in the previous section offer not only historical and cultural perspectives on stellar phenomena but also open up new pathways for engagement, education, and research. These stories—such as the Boorong record of Eta Carinae or the pulsating nature of Betelgeuse in Aboriginal oral traditions—demonstrate that traditional knowledge systems can complement and even guide scientific inquiry. For example, oral accounts describing the sudden appearance or disappearance of stars have inspired researchers to investigate them as potential records of supernovae or other transient events \citep{2018iau3.book...17H,2021acas.book..223N}. This suggests that Indigenous knowledge may assist in hypothesis generation within astrophysics, especially in reconstructing historical stellar activity.

In education, these narratives enhance student engagement by linking abstract concepts to vivid cultural stories. A tale like that of Nyeeruna’s fading fire magic, tied to Betelgeuse’s variability, makes the astrophysics of pulsating red giants more accessible and memorable for learners of all backgrounds. Initiatives such as Native Skywatchers and Indigenous astronomy camps in Australia and the U.S. have shown that culturally grounded content improves participation and curiosity among Indigenous youth and fosters respect among non-Indigenous students \citep{o2021watching,2020arXiv200805266L}. Experience from public presentations suggests that pairing narratives with astrophysical concepts can capture attention and encourage audiences to see familiar stars in new ways, highlighting the pedagogical value of Indigenous perspectives when applied to stellar phenomena.

Public outreach likewise benefits from these integrations. Planetarium shows that include stories like the Emu in the Sky (a dark nebula constellation) or the Māori tale of Matariki (the Pleiades) not only localize astronomy but also foster intergenerational learning and cross-cultural respect \citep{2025IAUS..384..379C}. These events often bring elders, families, and scientists together—blending oral storytelling with telescope observation in ways that resonate deeply with diverse communities. Observations from outreach settings indicate that when Indigenous narratives are introduced, audience engagement often shifts toward comparative questioning—how a cultural account aligns or diverges from astrophysical models—demonstrating that these integrations can actively deepen inquiry rather than remain purely illustrative.

To ensure that such integration is ethical and sustainable, collaborations must be rooted in Indigenous-led consent and governance. Successful models define content control, review processes, and benefit-sharing, such as those used in museum and planetarium programs, which allow Indigenous communities to curate narratives and retain authority over how their knowledge is shared \citep{2020arXiv200805266L}. Programs like ASKAP’s partnership with the Wajarri Yamatji people also illustrate how scientific infrastructure projects can be designed with cultural respect from the outset, as the telescope sits on their Country under an agreement that includes naming rights, heritage protections, and economic opportunities \citep{noon2024indigenous}.

The approach known as “Two-Eyed Seeing” encapsulates the strength of these collaborations: combining Indigenous and Western perspectives for a fuller understanding of the cosmos \citep{hamacher2023research}. Applied to stellar astrophysics, it encourages co-developed interpretations—for instance, understanding Antares both as Waiyungari, the glowing youth, and as a red supergiant approaching core collapse.

\section*{Acknowledgements}

I acknowledge that I work at the University of Texas at Austin, an institution located on land that has long been inhabited by Indigenous peoples. The Tonkawa lived in central Texas, and the Comanche and Apache also moved through and lived in this region. Today, Austin is home to Indigenous people from many different tribal nations. 

This work was supported by a travel grant from the International Astronomical Union for participation in IAU Symposium 399.



\bibliography{Sample}{}
\bibliographystyle{apalike}


\end{document}